\documentclass[twoside,english,sort&compress]{iopart}
\usepackage[T1]{fontenc}
\usepackage{geometry}
\usepackage{amsbsy}
\usepackage{amstext}
\usepackage{stmaryrd}
\usepackage{graphicx}
\usepackage{esint}
\usepackage[numbers]{natbib}
\usepackage{xcolor}

 \usepackage[normalem]{ulem} 
 \usepackage{soul} 

\makeatletter
\usepackage{iopams}
\usepackage{setstack}

\newcommand{\eqref}[1]{(\ref{#1})}

\makeatother

 \definecolor{red}{rgb}{1.0,0.0,0.0}
 \definecolor{gre}{rgb}{0.0,1.0,0.0}
 \definecolor{blu}{rgb}{0.0,0.0,1.0}

\usepackage{babel}
\begin{document}
\title[Pressure-Gradient-Driven Current Singularity]{Structure of Pressure-Gradient-Driven Current Singularity in Ideal
Magnetohydrodynamic Equilibrium}
\author{Yi-Min Huang}
\address{Department of Astrophysical Sciences and Princeton Plasma Physics
Laboratory, New Jersey 08543, USA}
\ead{yiminh@princeton.edu}

\author{Yao Zhou}
\address{School of Physics and Astronomy, Institute of Natural Sciences, and
MOE-LSC, Shanghai Jiao Tong University, Shanghai 200240, China }

\author{Joaquim Loizu}
\address{École Polytechnique Fédérale de Lausanne, Swiss Plasma Center, CH-1015
Lausanne, Switzerland }

\author{Stuart Hudson}
\address{Princeton Plasma Physics Laboratory, New Jersey 08543, USA}

\author{Amitava Bhattacharjee}
\address{Department of Astrophysical Sciences and Princeton Plasma Physics
Laboratory, New Jersey 08543, USA}
\begin{abstract}
Singular currents typically appear on rational surfaces in non-axisymmetric ideal magnetohydrodynamic equilibria with a continuum of nested flux surfaces and a continuous rotational transform. These currents have two components: a surface current (Dirac $\delta$-function in flux surface labeling) that prevents the formation of magnetic islands and an algebraically divergent Pfirsch--Schl\"uter current density when a pressure gradient is present across the rational surface. At flux surfaces adjacent to the rational surface, the traditional treatment gives the Pfirsch--Schl\"uter current density scaling as $J\sim1/\Delta\iota$, where $\Delta\iota$ is the difference of the rotational transform relative to the rational surface. If the distance $s$ between flux surfaces is proportional to $\Delta\iota$, the scaling relation $J\sim1/\Delta\iota\sim1/s$ will lead to a paradox that the Pfirsch--Schl\"uter current is not integrable. In this work, we investigate this issue by considering the pressure-gradient-driven singular current in the Hahm\textendash Kulsrud\textendash Taylor problem, which is a prototype for singular currents arising from resonant magnetic perturbations. We show that not only the Pfirsch--Schl\"uter current density but also the diamagnetic current density are divergent as $\sim1/\Delta\iota$. However, due to the formation of a Dirac $\delta$-function current sheet at the rational surface, the neighboring flux surfaces are strongly packed with $s\sim(\Delta\iota)^{2}$. Consequently, the singular current density $J\sim1/\sqrt{s}$, making the total current finite, thus resolving the paradox. Furthermore, the strong packing of flux surfaces causes a steepening of the pressure gradient near the rational surface, with $\nabla p \sim dp/ds \sim 1/\sqrt{s}$. In a general non-axisymmetric magnetohydrodynamic equilibrium, contrary to Grad's conjecture that the pressure profile is flat around densely distributed rational surfaces, our result suggests a pressure profile that densely steepens around them.

\end{abstract}
\noindent{\it Keywords\/}: {}
\submitto{\PPCF }
\maketitle

\section{Introduction}

Magnetic field configurations with nested flux surfaces are desirable
for plasma confinement fusion devices, due to the much faster motions
of charged particles along the magnetic field lines than in transverse
directions. Fusion devices with nested flux surfaces effectively insulate
hot plasmas from device walls, facilitating good confinement. Axisymmetric fusion devices such as tokamaks can
have a continuum of nested flux surfaces. However, for devices that
are inherently three-dimensional (3D), such as stellarators \citep{Spitzer1958,Boozer1998,Helander2014,Boozer2015a}, the existence of a continuum of nested flux surfaces is not guaranteed. 

Nevertheless, many existing magnetohydrodynamic (MHD) equilibrium
solvers for stellarators, e.g., VMEC \citep{HirshmanW1983}, NSTAB \citep{Taylor1994,Garabedian2002},
and DESC \citep{DudtK2020}, assume a continuum of nested flux surfaces
as a point of departure. These solvers seek equilibria satisfying
the MHD force balance equation 
\begin{equation}
-\nabla p+\boldsymbol{J}\times\boldsymbol{B}=0\label{eq:Force}
\end{equation}
either by minimizing the MHD potential energy (i.e., the sum of the magnetic and the thermal energies) or by directly imposing
the force balance as a constraint. 
Here, $p$, $\boldsymbol{B}$, and $\boldsymbol{J}=\nabla\times\boldsymbol{B}$
are standard notations for the plasma pressure, the magnetic field, and the current
density. 

Toroidal 3D MHD equilibria with a continuum of nested flux surfaces
can potentially give rise to current singularities at rational surfaces,
where magnetic field lines form closed loops rather than fill the
flux surfaces ergodically. To understand the issue, note that the
force balance equation (\ref{eq:Force}) implies that the component of the
current density perpendicular to the magnetic field is given by
\begin{equation}
\boldsymbol{J}_{\perp}=\frac{\boldsymbol{B}\times\nabla p}{B^{2}}.\label{eq:J_perp}
\end{equation}
By writing 
\begin{equation}
\boldsymbol{J}=\frac{J_{\parallel}}{B}\boldsymbol{B}+\boldsymbol{J}_{\perp},\label{eq:decompose}
\end{equation}
the condition $\nabla\cdot\boldsymbol{J}=0$ implies
\begin{equation}
\boldsymbol{B}\cdot\nabla\frac{J_{\parallel}}{B}=-\nabla\cdot\boldsymbol{J}_{\perp}=-\left(\boldsymbol{B}\times\nabla p\right)\cdot\nabla\frac{1}{B^{2}},\label{eq:magnetic_de}
\end{equation}
and in general $\nabla\cdot\boldsymbol{J}_{\perp}\neq0$ if $\nabla p\neq0$.
In a straight-field-line coordinate system $(\Psi,\theta,\varphi)$,
the magnetic field is given by
\[
\boldsymbol{B}=\nabla\Psi\times\left(\nabla\theta-\iota\nabla\varphi\right),
\]
where $\Psi$ is the toroidal flux function, $\theta$ is a poloidal
angle, $\varphi$ is a toroidal angle, and $\iota=\iota\left(\Psi\right)$
is the rotational transform. In these coordinates, the magnetic differential operator $\boldsymbol{B}\cdot\nabla$
is given by 
\begin{equation}
\boldsymbol{B}\cdot\nabla f=\frac{1}{\mathcal{J}}\left(\frac{\partial f}{\partial\varphi}+\iota\frac{\partial f}{\partial\theta}\right),\label{eq:Bdg}
\end{equation}
where $\mathcal{J}$ is the Jacobian of the coordinates. By using
Fourier representations 
\begin{equation}
J_{\parallel}/B=\sum_{m,n}u_{mn}\left(\Psi\right)\exp\left[i\left(m\theta-n\varphi\right)\right]\label{eq:Fourier1}
\end{equation}
and 
\begin{equation}
\mathcal{J}\nabla\cdot\boldsymbol{J}_{\perp}=\sum_{m,n}h_{mn}\left(\Psi\right)\exp\left[i\left(m\theta-n\varphi\right)\right],\label{eq:Fourier2}
\end{equation}
the magnetic differential equation (\ref{eq:magnetic_de}) can be
written as
\begin{equation}
\sum_{m,n}i\left(\iota\left(\Psi\right)m-n\right)u_{mn}\left(\Psi\right)\exp\left[i\left(m\theta-n\varphi\right)\right]=-\sum_{m,n}h_{mn}\left(\Psi\right)\exp\left[i\left(m\theta-n\varphi\right)\right].\label{eq:magnetic_de1}
\end{equation}
Therefore, $J_{\parallel}/B$ can be expressed as
\begin{equation}
J_{\parallel}/B=\sum_{m,n\neq0,0}\left[i\frac{h_{mn}\left(\Psi\right)}{\iota\left(\Psi\right)m-n}+\Delta_{mn}\delta\text{\ensuremath{\left(\iota\left(\Psi\right)m-n\right)}}\right]\exp\left[i\left(m\theta-n\varphi\right)\right]+u_{00}\left(\Psi\right),\label{eq:J_parallel-1}
\end{equation}
where $u_{00}$ is a flux function and $\Delta_{mn}$ are coefficients
to be determined \citep{CaryM1985, HegnaB1989}. 

Equation (\ref{eq:J_parallel-1}) exhibits two types of singularities
at rational surfaces, where the rotational transform $\iota$ are
rational numbers. Let $x=\iota\left(\Psi\right)m-n$. The first term
in Eq.~(\ref{eq:J_parallel-1}), known as the Pfirsch--Schl\"uter
current, has a $1/x$-type singularity at the rational surface with
$\iota=n/m$. The second term is a Dirac $\delta$-function singularity,
which is permitted because $x\delta(x)=0$. 

The physical quantity associated with the current density is the total
current through a surface. The Dirac $\delta$-function type singularities
are sheet currents at rational surfaces. Even though the current density
becomes infinite, the total current is finite because the Dirac $\delta$
function is integrable. The $1/x$-type singularities of the Pfirsch--Schl\"uter
current pose a more serious problem. If the distances between the
rational surface and neighboring flux surfaces are proportional to
$x$, then the total current through a constant-$\varphi$ surface element enclosed by
$\theta$, $\theta+d\theta$, $x=\epsilon_{1}$, and $x=\epsilon_{2}$
is proportional to $\int_{\epsilon_{1}}^{\epsilon_{2}}dx'/x'$, which
diverges logarithmically when $\epsilon_{1}\to0$. For this reason,
the $1/x$-type singularities are considered unphysical and should
be avoided. 

The $1/x$-type singularities can be eliminated if the plasma pressure is globally flat, but that is not of interest to magnetic confinement fusion. Grad noted that magnetically confined MHD equilibrium solutions may be constructed by considering pressure profiles that are flat in a small neighborhood of each rational surface. However, if the rotational transform $\iota\left(\psi\right)$ is continuous, the magnetic differential equation (\ref{eq:magnetic_de}) is densely singular and there will be infinitely many flat-pressure regions throughout the entire plasma volume. Grad described such a pressure distribution as pathological \citep{Grad1967}. Along this line of thought, Hudson and Kraus constructed fractal-like pressure profiles with gradients localized to where the rotational transforms are sufficiently irrational, i.e., those that satisfy a Diophantine condition  \citep{HudsonK2017,KrausH2017}. Alternatively, one may allow the rotational transform $\iota$ to have discontinuities such that rational surfaces do not exist.

The central difficulty associated with the $1/x$-type singularities is that they appear to give rise to divergent currents. However, this conclusion is obtained through heuristic arguments
as we have discussed above rather than through a rigorous mathematical
proof. The primary objective of this work is to reassess this conclusion
through a simple model problem \textemdash{} the Hahm--Kulsrud--Taylor
(HKT) problem \citep{HahmK1985, DewarBKW2013}. The HKT problem was originally posed
as a forced reconnection problem, but in the ideal MHD limit, it also
serves as a prototype for current singularity formation driven by
resonant magnetic perturbations \citep{ZhouHQB2016}. The HKT problem is amenable
to analytic solutions \citep{ZhouHRQB2019,HuangHLZB2022} and has been studied with various numerical codes including a Grad\textendash Shafranov (GS) solver \citep{HuangBZ2009},
a fully Lagrangian code \citep{ZhouQBB2014}, and the Stepped Pressure
Equilibrium Code (SPEC)\citep{HudsonDDHMNL2012} for the case with
$p=0$ \citep{ZhouHQB2016,ZhouHQB2018,ZhouHRQB2019,HuangHLZB2022}.
In this study, we extend the established analytic and numerical solutions
to include a non-vanishing pressure gradient. 

This paper is organized as follows. Section \ref{sec:Grad-Shafranov-solutions}
lays out the Grad\textendash Shafranov formulation of the HKT problem
and provides an asymptotic analytic solution. Section \ref{sec:Numerical-Solutions}
compares the analytic solution with numerical solutions. We conclude
and discuss future outlook in Section \ref{sec:Discussions-and-Conclusion}.

\section{Grad\textendash Shafranov Solution of the Hahm\textendash Kulsrud\textendash Taylor
Problem with Pressure Gradient \label{sec:Grad-Shafranov-solutions}}

\begin{figure}
\begin{centering}
\includegraphics[width=0.8\columnwidth]{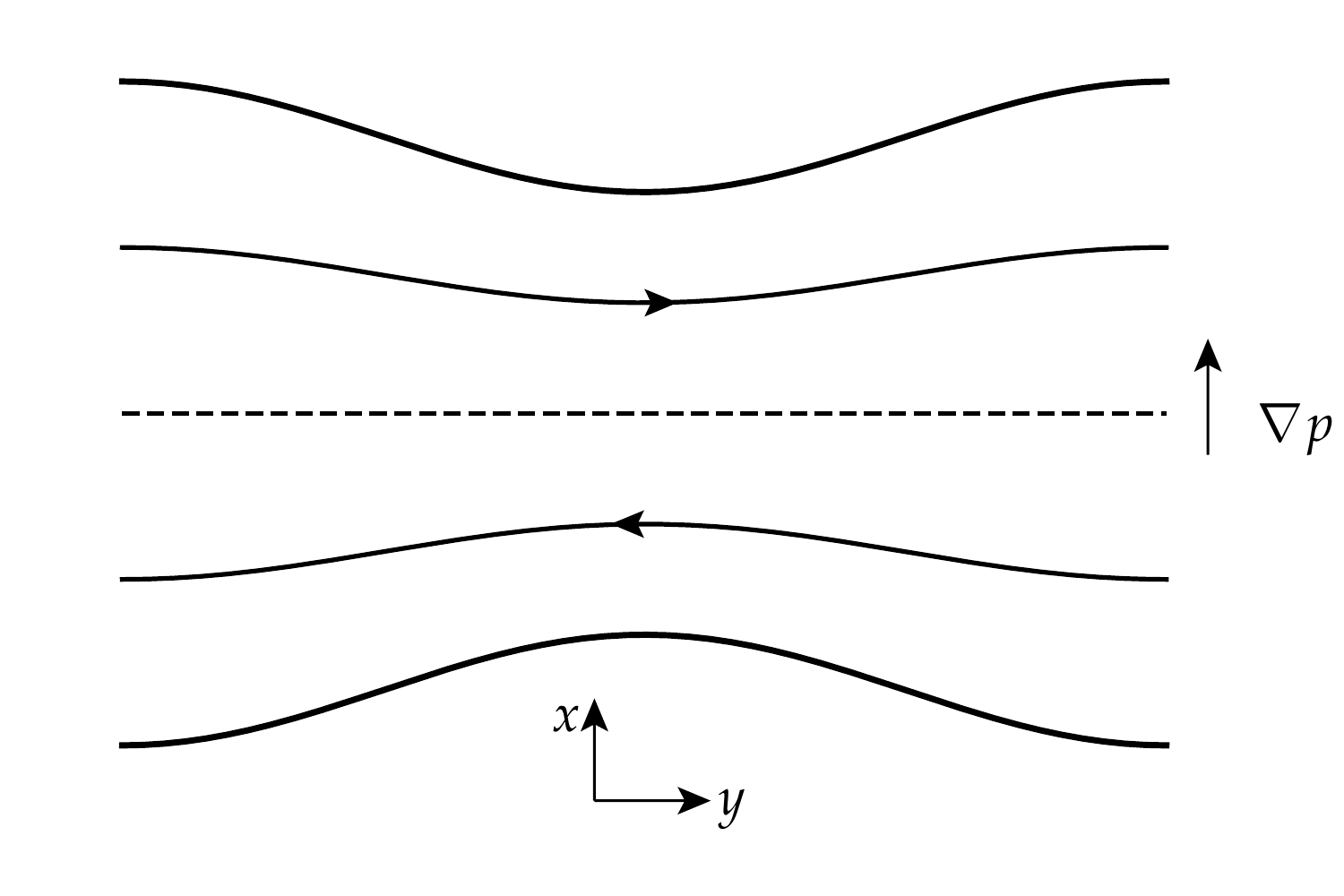}
\par\end{centering}
\caption{The Hahm--Kulsrud--Taylor problem with a pressure gradient. The in-plane
components of the magnetic field reverse directions at the mid-plane
(the dashed line). The upper and lower boundaries are shaped by mirror-symmetric
sinusoidal perturbations. In response to the perturbation, a $\delta$-function
singular current sheet develops at the mid-plane, which resonates
with the boundary perturbations. The pressure gradient of the plasma
drives the Pfirsch\textendash Schl\"uter current density, which diverges
algebraically near the resonant surface. \label{fig:A-sketch-of-HKT}}
\end{figure}

\subsection{Grad\textendash Shafranov Formulation for the Hahm\textendash Kulsrud\textendash Taylor
Problem}

The Hahm\textendash Kulsrud\textendash Taylor (HKT) problem, illustrated
in Fig.~\ref{fig:A-sketch-of-HKT}, has a magnetized plasma enclosed
by two conducting walls in slab geometry. Before the conducting
walls are perturbed, the initial magnetic field is a smooth function
of space in the domain with $-a\le x\le a$ and $0\le y\le L$. The
in-plane component points in the $y$ direction with $B_{y0}=x$;
together with a non-uniform $B_{z0}$ component and the pressure $p_{0}$,
the system is in force balance; i.e., the equation 
\begin{equation}
\frac{B_{y0}^{2}+B_{z0}^{2}}{2}+p_{0}=\text{const}\label{eq:1Deq}
\end{equation}
is satisfied. Assuming a periodic boundary condition along the $y$
direction, we then impose a sinusoidal perturbation with an up-down
symmetry that displaces the conducting walls at $x=\pm a$ to $x=\pm\left(a+\delta\cos\left(ky\right)\right)$,
where $k=2\pi/L$ is the wavenumber. As the system evolves to a new
equilibrium subject to the ideal MHD constraint, current singularities,
including a Dirac $\delta$-function current sheet and a pressure-gradient-driven
current singularity, will develop at and around the resonant surface $x=0$ (the dashed
line in Figure \ref{fig:A-sketch-of-HKT}).

In response to the boundary perturbation, the flux surfaces adjust
their geometries while preserving the magnetic fluxes between them in
accordance with the ideal MHD constraints. The new MHD equilibrium
satisfies the Grad\textendash Shafranov equation in Cartesian
geometry
\begin{equation}
\nabla^{2}\psi=-\frac{dP}{d\psi},\label{eq:GS}
\end{equation}
where 
\begin{equation}
P=p+\frac{B_{z}^{2}}{2}.\label{eq:total_pressure}
\end{equation}
Here, the magnetic field in Cartesian geometry, with
$z$ being the direction of translational symmetry, is expressed in
terms of the poloidal (in-plane) flux function $\psi$ as 
\begin{equation}
\boldsymbol{B}=B_{z}\boldsymbol{\hat{z}}+\boldsymbol{\hat{z}}\times\nabla\psi.\label{eq:B}
\end{equation}
In an equilibrium state, both the out-of-plane component $B_{z}$
and the plasma pressure $p$ are functions of $\psi$. 

Because the magnetic fluxes are conserved, the magnetic field is determined
by the geometry of flux surfaces. To describe the geometry of
flux surfaces, we first need to choose a variable to label them. Let
$x_{0}$ be the positions of flux surfaces along the $x$ direction
before the boundary perturbation is imposed. The initial condition
$B_{y0}=x_{0}$ yields the poloidal (in-plane) magnetic flux function
$\psi\left(x_{0}\right)=x_{0}^{2}/2$. While it is customary in toroidal
confinement devices to label flux surfaces by the magnetic flux function,
for the HKT problem it is convenient to label the flux surfaces with
their initial positions $x_{0}$ because a single value of $\psi$
can correspond to more than one flux surface. 

With this labeling, the geometry of flux surfaces is described
by a mapping from $\left(x_{0},y\right)$ to $\left(x,y\right)$ via
a function $x\left(x_{0},y\right)$. Using the chain rule, we can
express the partial derivatives with respect to the Cartesian coordinates
in terms of the partial derivatives to the coordinates $\left(x_{0},y\right)$:
\begin{equation}
\left(\frac{\partial}{\partial x}\right)_{y}=\frac{1}{\partial x/\partial x_{0}}\frac{\partial}{\partial x_{0}},\label{eq:partial_x}
\end{equation}
\begin{equation}
\left(\frac{\partial}{\partial y}\right)_{x}=\frac{\partial}{\partial y}-\frac{\partial x/\partial y}{\partial x/\partial x_{0}}\frac{\partial}{\partial x_{0}}.\label{eq:partial_y}
\end{equation}
Here, the subscripts of the partial derivatives on the left-hand side
indicate the coordinates that are held fixed; the partial derivatives
on the right-hand side are with respect to the $\left(x_{0},y\right)$
coordinates. Hereafter, partial derivatives are taken to be with respect
to the $\left(x_{0},y\right)$ coordinates by default, unless otherwise
indicated by the subscripts. 

Using these relations for partial derivatives, the Cartesian components
of the in-plane magnetic field are given by
\begin{equation}
B_{x}=-\left(\frac{\partial\psi}{\partial y}\right)_{x}=\frac{\partial x/\partial y}{\partial x/\partial x_{0}}\frac{d\psi}{dx_{0}}\label{eq:Bx}
\end{equation}
and 
\begin{equation}
B_{y}=\left(\frac{\partial\psi}{\partial x}\right)_{y}=\frac{1}{\partial x/\partial x_{0}}\frac{d\psi}{dx_{0}}.\label{eq:By}
\end{equation}

The out-of-plane component $B_{z}$ is determined by the conservation
of the out-of-plane magnetic flux as 
\begin{equation}
B_{z}\left(x_{0}\right)=\frac{B_{z0}\left(x_{0}\right)}{\left\langle \frac{\partial x}{\partial x_{0}}\right\rangle }.\label{eq:Bz}
\end{equation}
Here, $B_{z0}$ is the initial $z$-component of the magnetic field
and the flux surface average $\left\langle f\right\rangle $ is defined
as 
\begin{equation}
\left\langle f\right\rangle \equiv\frac{1}{L}\int_{0}^{L}f\left(x_{0},y\right)dy\label{eq:ave}
\end{equation}
for an arbitrary function $f\left(x_{0},y\right)$, where $L$ is
the period of the system along the $y$ direction. Likewise, the pressure
is determined by the adiabatic energy equation as
\begin{equation}
p\left(x_{0}\right)=\frac{p_{0}\left(x_{0}\right)}{\left\langle \frac{\partial x}{\partial x_{0}}\right\rangle ^{\gamma}},\label{eq:pressure}
\end{equation}
where $p_{0}\left(x_{0}\right)$ is the initial pressure profile and
$\gamma$ is the adiabatic index.  

From the relation $\boldsymbol{J=}\nabla\times\boldsymbol{B}$, the
out-of-plane component of the current density is given by
\begin{equation}
J_{z}=\nabla^{2}\psi=\left(\frac{d\psi}{dx_{0}}\right)^{-1}\frac{\partial}{\partial x_{0}}\left(\frac{B_{x}^{2}+B_{y}^{2}}{2}\right)-\frac{\partial B_{x}}{\partial y}\label{eq:Jz}
\end{equation}
and the poloidal (in-plane) component $\boldsymbol{J}_{p}$ is 
\begin{equation}
\boldsymbol{J}_{p}=\frac{dB_{z}}{d\psi}\nabla\psi\times\boldsymbol{\hat{z}}=-\frac{dB_{z}}{d\psi}\boldsymbol{B}_{p},\label{eq:Jp}
\end{equation}
where $\boldsymbol{B}_{p}=\boldsymbol{\hat{z}}\times\nabla\psi$ is the poloidal (in-plane) component of the magnetic field. Substituting Eq.~(\ref{eq:Jz}) in Eq.~(\ref{eq:GS}), the GS equation
can be written as 
\begin{equation}
-\frac{\partial}{\partial x_{0}}\left(\frac{B_{x}^{2}+B_{y}^{2}}{2}+P\right)+\frac{\partial B_{x}}{\partial y}\frac{d\psi}{dx_{0}}=0.\label{eq:GS1}
\end{equation}

\subsection{Outer-Region Solution}

For a small boundary perturbation, we may linearize the GS equation
in terms of the displacements of the flux surfaces along the $x$ direction  $\xi\left(x_{0},y\right)\equiv x\left(x_{0},y\right)-x_{0}$.
To the leading order of $\xi$, the magnetic field components are
\begin{equation}
B_{x}\simeq\frac{\partial\xi}{\partial y}\frac{d\psi}{dx_{0}},\label{eq:Bx1}
\end{equation}
\begin{equation}
B_{y}\simeq\left(1-\partial\xi/\partial x_{0}\right)\frac{d\psi}{dx_{0}},\label{eq:By1}
\end{equation}
and 
\begin{equation}
B_{z}\simeq B_{z0}\left(x_{0}\right)\left(1-\left\langle \frac{\partial\xi}{\partial x_{0}}\right\rangle \right).\label{eq:Bz1}
\end{equation}
The linearized pressure is
\begin{equation}
p\simeq p_{0}\left(x_{0}\right)\left(1-\gamma\left\langle \frac{\partial\xi}{\partial x_{0}}\right\rangle \right).\label{eq:pressure1}
\end{equation}
Using Eqs. (\ref{eq:Bx1}\textendash \ref{eq:pressure1}) and the
relation $\psi=x_0^2/2$ in Eq.~(\ref{eq:GS1}), the linearized GS equation now reads
\begin{equation}
\frac{\partial}{\partial x_{0}}\left(x_{0}^{2}\frac{\partial\xi}{\partial x_{0}}+\left(B_{z0}^{2}+p_{0}\gamma\right)\left\langle \frac{\partial\xi}{\partial x_{0}}\right\rangle \right)+x_{0}^{2}\frac{\partial^{2}\xi}{\partial y^{2}}=0.\label{eq:GS-linear}
\end{equation}

If we adopt the ansatz $\xi=\bar{\xi}(x_{0})\cos(ky)$, then $\left\langle \partial\xi/\partial x_{0}\right\rangle =0$
and the linearized GS equation reduces to
\begin{equation}
\frac{d^{2}}{dx_{0}^{2}}\left(x_{0}\bar{\xi}\right)-k^{2}x_{0}\bar{\xi}=0.\label{eq:GS-linear_HKT}
\end{equation}
The general solution of Eq.~(\ref{eq:GS-linear_HKT}) is a linear
superposition of two independent solutions
\begin{equation}
\bar{\xi}=c_{1}\frac{\sinh\left(k\left|x_{0}\right|\right)}{x_{0}}+c_{2}\frac{\cosh\left(kx_{0}\right)}{x_{0}},\label{eq:linear_sol}
\end{equation}
and the boundary condition $\bar{\xi}(\pm a)=\pm\delta$ requires
\begin{equation}
\delta=\frac{c_{1}\sinh\left(ka\right)+c_{2}\cosh\left(ka\right)}{a}.\label{eq:BC-2}
\end{equation}

The independent solution $\cosh\left(kx_{0}\right)/x_{0}$ in Eq.~(\ref{eq:linear_sol})
diverges at $x_{0}=0$. If we insist that the linear solution is valid
everywhere, then we must set $c_{2}=0$; the boundary condition
(\ref{eq:BC-2}) then determines the coefficient $c_{1}=a\delta/\sinh\left(ka\right)$.
However, this solution is not physically tenable. Because $\lim_{x_{0}\to0}\sinh\left(k\left|x_{0}\right|\right)/x_{0}=k \, \text{sgn}(x_0)$,
the geometry of the flux surfaces is approximately given by $x\simeq x_{0}+\,\text{sgn}(x_0)\left(ka\delta/\sinh\left(ka\right)\right)\cos\left(ky\right)$
in the vicinity of $x_{0}=0$. Hence, the flux surfaces with $\left|x_{0}\right|\lesssim ka\delta/\sinh\left(ka\right)$
overlap, causing a condition that is physically not permitted. We
conclude that the linear solution is not valid within an inner region
with $\left|x_{0}\right|\lesssim O\left(ka\delta/\sinh\left(ka\right)\right)$
and a nonlinear solution must be sought.

\subsection{Inner-Layer Solution}

A general nonlinear solution for the inner layer near a resonant surface
was first developed by Rosenbluth, Dagazian, and Rutherford (hereafter
RDR) for the bifurcated equilibrium after an ideal internal kink instability \citep{RosenbluthDR1973}.
The RDR solution was later adapted to the HKT problem with $p=0$ \citep{ZhouHRQB2019,HuangHLZB2022}.
Here, we generalize previous approaches to incorporate the pressure-gradient
effects. 

Because the inner region is a thin layer, we assume the conditions
$\left|\partial x/\partial y\right|\ll1$ and $\left|\partial/\partial y\right|$$\ll\left|\partial/\partial x_{0}\right|$
are satisfied. Under these assumptions, the dominant balance of the
GS equation (\ref{eq:GS1}) is given by

\begin{equation}
\frac{\partial}{\partial x_{0}}\left(\frac{B_{y}^{2}}{2}+P\left(x_{0}\right)\right)=0.\label{eq:GS_inner}
\end{equation}
Integrating Eq.~(\ref{eq:GS_inner}) yields 
\begin{equation}
B_{y}=\frac{1}{\partial x/\partial x_{0}}\frac{d\psi}{dx_{0}}=\text{sgn}\left(\frac{d\psi}{dx_{0}}\right)\sqrt{f(x_{0})+g(y)},\label{eq:Inner_sol}
\end{equation}
where 
\begin{equation}
f\left(x_{0}\right)=-2P\left(x_{0}\right)+\text{const}\label{eq:constraint}
\end{equation}
and $g(y)$ is an arbitrary function that will be determined later
by asymptotic matching of the inner-layer solution to the outer-region solution. The $\text{sgn}\left(d\psi/dx_{0}\right)$
factor comes from the requirement that $\partial x/\partial x_{0}>0$
must be satisfied to avoid overlapping flux surfaces. Without loss
of generality, we are free to set $f(0)=0$, and the tangential discontinuity
of $B_{y}$ at $x_{0}=0$ is then given by
\begin{equation}
\left.B_{y}\right|_{0^{\pm}}=\pm\sqrt{g(y)}.\label{eq:tangential_discontinuity}
\end{equation}
This tangential discontinuity corresponds to a Dirac $\delta$-function
singularity in the current density. 

Using the flux function $\psi=x_{0}^{2}/2$ for the HKT problem and
integrating Eq. (\ref{eq:Inner_sol}) again yields the inner-layer
solution of RDR
\begin{equation}
x_{\text{RDR}}\left(x_{0},y\right)=h(y)+\int_{0}^{x_{0}}\frac{\left|x'\right|}{\sqrt{f(x')+g(y)}}dx',\label{eq:RDR}
\end{equation}
where the function $h(y)$, which describes the geometry of the resonant
surface, is yet to be determined. 

\subsection{Incompressibility Constraint}


The functions $f\left(x_{0}\right)$ and $g\left(y\right)$
are not independent, but are implicitly related through Eq.~(\ref{eq:constraint}) as we will see below. First, we determine $B_{z}$ and $p$ in the inner layer by substituting
$x_{\text{RDR}}\left(x_{0},y\right)$ for $x\left(x_{0},y\right)$
in Eqs. (\ref{eq:Bz}) and (\ref{eq:pressure}). Then, applying the obtained
$B_{z}$ and $p$ in Eq.~(\ref{eq:constraint}), the resulting relation
\begin{eqnarray}
f\left(x_{0}\right) & = & -\frac{B_{z0}\left(x_{0}\right)^{2}}{|x_{0}|^{2}}\left\langle \left(f(x_{0})+g(y)\right)^{-1/2}\right\rangle ^{-2}\nonumber \\
 &  & -2\frac{p_{0}\left(x_{0}\right)}{|x_{0}|^{\gamma}}\left\langle \left(f(x_{0})+g(y)\right)^{-1/2}\right\rangle ^{-\gamma}+\text{const}\label{eq:constraint1}
\end{eqnarray}
gives a relation between $f\left(x_{0}\right)$ and $g\left(y\right)$.

We now show that Eq.~(\ref{eq:constraint1}) is approximately equivalent to the incompressible constraint in the strong-guide-field limit with $B_{p}^{2}\sim p\ll B_{z}^{2}$. From Eq.~(\ref{eq:constraint1}), the differentials $df$ and $dx_{0}$ are related by 
\begin{eqnarray}
df & = & \frac{2B_{z0}{}^{2}}{|x_{0}|^{2}}\left\langle \left(f+g\right)^{-1/2}\right\rangle ^{-2}\left(\frac{1}{x_{0}}dx_{0}-\frac{1}{2}\left\langle \left(f+g\right)^{-1/2}\right\rangle ^{-1}\left\langle \left(f+g\right)^{-3/2}\right\rangle df\right)\nonumber \\
 &  & +\frac{2\left(p_{0}'+B_{y0}B_{y0}'\right)}{|x_{0}|^{2}}\left\langle \left(f+g\right)^{-1/2}\right\rangle ^{-2}dx_{0}\nonumber \\
 &  & +\left(2\gamma\frac{p_{0}}{|x_{0}|^{\gamma}x_{0}}-2\frac{p_{0}'}{|x_{0}|^{\gamma}}\right)\left\langle \left(f+g\right)^{-1/2}\right\rangle ^{-\gamma}dx_{0}\nonumber \\
 &  & -\gamma\frac{p_{0}}{|x_{0}|^{\gamma}}\left\langle \left(f+g\right)^{-1/2}\right\rangle ^{-\gamma-1}\left\langle \left(f+g\right)^{-3/2}\right\rangle df,\label{eq:df1}
\end{eqnarray}
where we have used the relation $p_{0}'+B_{z0}B_{z0}'+B_{y0}B_{y0}'=0$
to eliminate $B_{z0}'$. In the strong-guide-field limit,
the first term in the right-hand side of Eq.~(\ref{eq:df1}) is the
dominant term. As the leading order approximation, we may ignore other
terms in the equation, resulting in the relation
\begin{equation}
\frac{1}{x_{0}}dx_{0}\simeq\frac{1}{2}\left\langle \left(f+g\right)^{-1/2}\right\rangle ^{-1}\left\langle \left(f+g\right)^{-3/2}\right\rangle df.\label{eq:leading_order_eq}
\end{equation}
Integrating Eq.~$(\ref{eq:leading_order_eq})$ yields
\begin{equation}
\left|x_{0}\right|\simeq C\left\langle \left(f+g\right)^{-1/2}\right\rangle ^{-1},\label{eq:leading_order-sol}
\end{equation}
where $C$ is a constant. Finally, we apply Eq.~(\ref{eq:leading_order-sol})
in Eq.~(\ref{eq:RDR}) and obtain 
\begin{equation}
\left\langle \frac{\partial x_{\text{RDR}}}{\partial x_{0}}\right\rangle =\left|x_{0}\right|\left\langle \left(f+g\right)^{-1/2}\right\rangle \simeq C.\label{eq:compression_ratio}
\end{equation}

Equation (\ref{eq:compression_ratio}) has a simple physical meaning:
The plasma in the inner layer is uniformly compressed or expanded,
and the constant $C$ is the compression ratio. Because the inner-layer
solution is to be matched to the outer-region solution and the latter
is not compressive (up to the linear approximation), the compression ratio must assume the value $C=1$;
that is, the plasma approximately satisfies the incompressibility constraint.
It should be noted, however, that the incompressibility constraint will
not be valid if the boundary perturbation has an $m=0$ Fourier component. In this case, the constant $C$ needs to be determined by the asymptotic matching as well.

\subsection{Asymptotic Matching}

The inner-layer solution must be asymptotically matched with the outer-region
solution to have a complete solution. For simplicity, we will assume
the strong-guide-field limit and replace the relation (\ref{eq:constraint1})
by the incompressibility constraint 
\begin{equation}
\left|x_{0}\right|=\left\langle \left(f+g\right)^{-1/2}\right\rangle ^{-1}.\label{eq:incompressible}
\end{equation}
In this limit, to the leading order approximation, the effect of pressure gradient on the geometry of flux surfaces is negligible. Moreover,
because the pressure gradient also does not affect the outer-region
solution, the asymptotic matching procedure, which we will outline below, is identical to that for the case with $p=0$ \citep{ZhouHRQB2019}. 

The outer-region solution is expressed in terms of the displacements $\xi$ of flux surfaces. Therefore, to facilitate the matching, we express the flux surface displacement of the inner-layer solution as
\begin{equation}
\xi\left(x_{0},y\right)=h(y)+\int_{0}^{x_{0}}\left(\frac{\left|x'\right|}{\sqrt{f(x')+g(y)}}-1\right)dx'\label{eq:RDR_displacement}
\end{equation}
and examine its asymptotic behavior as $x_{0}\to\pm\infty$. 

From the incompressibility constraint (\ref{eq:incompressible}), we can infer that $f\to x_{0}^{2}$ as $x_{0}\to\pm\infty$. Hence, in
the asymptotic limit of $x_{0}\to\pm\infty$ we can split the integral
in Eq.~(\ref{eq:RDR_displacement}) into two parts, yielding
\begin{eqnarray}
\xi\left(x_{0},y\right) & \simeq & h(y)+\text{sgn}(x_{0})\int_{0}^{\infty}\left(\frac{x'}{\sqrt{f(x')+g(y)}}-1\right)dx'\nonumber \\
 &  & -\text{sgn}(x_{0})\int_{|x_{0}|}^{\infty}\left(\frac{x'}{\sqrt{x'^{2}+g(y)}}-1\right)dx'\nonumber \\
 & \simeq & h(y)+\text{sgn}(x_{0})\left[\int_{0}^{\infty}\left(\frac{x'}{\sqrt{f(x')+g(y)}}-1\right)dx'\right]+\frac{1}{2}\frac{g(y)}{x_{0}}.\label{eq:RDR_asymptotic}
\end{eqnarray}
This asymptotic behavior of the inner-layer solution as $x_{0}\to\pm\infty$
should match the behavior of the outer-region solution in the limit
of $x_{0}\to0^{\pm}$, which is
\begin{equation}
\xi(x_{0},y)\simeq\left[\text{sgn}(x_{0})c_{1}k+\frac{c_{2}}{x_{0}}\right]\cos(ky).\label{eq:outer_asymptotic}
\end{equation}
Matching Eqs. (\ref{eq:RDR_asymptotic}) and (\ref{eq:outer_asymptotic})
yields $h(y)=0$ and the following two conditions: 
\begin{equation}
\int_{0}^{\infty}\left(\frac{x'}{\sqrt{f(x')+g(y)}}-1\right)dx'=c_{1}k\cos(ky)\label{eq:matching1}
\end{equation}
and 
\begin{equation}
\frac{1}{2}\frac{g(y)}{x_{0}}\leftrightarrow\frac{c_{2}}{x_{0}}\cos(ky).\label{eq:matching2}
\end{equation}
Here, we use the notation $\leftrightarrow$ in the second condition
because, as we will see later, Eq.~(\ref{eq:matching2}) can not
be matched exactly. 

The matching condition (\ref{eq:matching1}) gives an integral equation
to determine the function $g(y)$ as follows: by eliminating $x'$
in favor of $f$ using the incompressibility constraint (\ref{eq:incompressible})
in Eq.~(\ref{eq:matching1}), we obtain 

\begin{eqnarray}
 & \int_{0}^{\infty}\left((f+g)^{-1/2}-\left\langle (f+g)^{-1/2}\right\rangle \right)\left\langle (f+g)^{-1/2}\right\rangle ^{-3}\left\langle \left(f+g\right)^{-3/2}\right\rangle df\nonumber \\
= & 2c_{1}k\cos(ky).\label{eq:RDR_integral}
\end{eqnarray}
However, analytically solving this equation to obtain $g(y)$ is a
daunting task. By studying the behavior of $g(y)$ around $y=0$ and
$y=\pi/k$, RDR suggested a function form $g(y)\propto\sin^{8}(ky/2)$
as an approximate solution. This suggestion was confirmed by Loizu
and Helander with a numerical solution of the integral equation \citep{LoizuH2017}.
Moreover, by fitting the function form with the numerical solution, they also determined the coefficient in front, yielding 
\begin{equation}
g(y)\simeq\frac{4c_{1}^{2}k^{2}}{3}\sin^{8}(ky/2).\label{eq:gy}
\end{equation}

Next, for the matching condition (\ref{eq:matching2}), it is clear
that $g(y)$ cannot be matched with $\cos(ky)$ exactly. Following
RDR, we expand $g(y)$ as a Fourier series
\begin{equation}
g(y)=\sum_{m=0}^{\infty}\Gamma_{m}\cos(mky)\label{eq:expansion}
\end{equation}
and only match the $m=1$ term; that gives
\begin{equation}
c_{2}=-\frac{7c_{1}^{2}k^{2}}{24}.\label{eq:c2_c1_relation}
\end{equation}

Finally, we use the relation (\ref{eq:c2_c1_relation}) to eliminate
$c_{2}$ in the boundary condition (\ref{eq:BC-2}), yielding an equation
for $c_{1}$:
\begin{equation}
7k^{2}\cosh(k/2)c_{1}^{2}-24\sinh(k/2)c_{1}+12\delta=0.\label{eq:c1_eq}
\end{equation}
The solution is
\begin{equation}
c_{1}=\frac{12\sinh(ka)-\sqrt{144\sinh^{2}(ka)-84\delta k^{2}\cosh(ka)}}{7k^{2}\cosh(ka)}.\label{eq:c1}
\end{equation}
Here, the sign for the square root in Eq.~(\ref{eq:c1}) is chosen such
that $c_{1}\simeq\delta/2\sinh(k/2)$ in the limit of a small perturbation. 

Now, we have obtained $c_{1}$, $c_{2}$, $h(y)$, and $g(y)$. We
can then solve Eq.~(\ref{eq:incompressible}) to obtain $f(x_{0})$,
which completes the necessary information to calculate the RDR solution,
Eq.~(\ref{eq:RDR}). Solving $f(x_{0})$ from Eq.~(\ref{eq:incompressible})
in general requires a numerical treatment. However, the leading order
behavior of $f(x_{0})$ in the limit of $\left|x_{0}\right|\to0$
can be obtained analytically as \citep{HuangHLZB2022}
\begin{equation}
f(x_{0})\simeq\left(c_{f}\left|x_{0}\right|\right)^{8/3},\label{eq:f_0}
\end{equation}
where the coefficient $c_{f}$ can be expressed in terms of the gamma
function $\Gamma$ \citep{AbramowitzS1972} as
\begin{equation}
c_{f}\equiv\frac{2(3/4)^{1/8}}{\pi^{3/2}}\Gamma(3/8)\Gamma(9/8)\left(c_{1}k\right)^{-1/4}\simeq0.7735\left(c_{1}k\right)^{-1/4}.\label{eq:cf}
\end{equation}

\subsection{Pressure-Gradient-Driven Current Singularity}

Our analysis thus far has shown that the effect of pressure gradient on the geometry of flux surfaces is negligible. We are now in a position to discuss the effect of pressure gradient on the current
density distribution. We are primarily interested in the current density
in the vicinity of the resonant surface, i.e., the inner-layer solution. 

In the strong-guide-field limit, the plasma is approximately incompressible;
therefore, from Eqs.~(\ref{eq:Bz}) and (\ref{eq:pressure}) we have
$p\left(x_{0}\right)\simeq p_{0}\left(x_{0}\right)$ and $B_{z}\left(x_{0}\right)\simeq B_{z0}\left(x_{0}\right)$
as the leading order approximation. To calculate the current density
requires $B_{z}'\left(x_{0}\right)$, but we cannot simply assume
$B_{z}'\left(x_{0}\right)\simeq B_{z0}'\left(x_{0}\right)$ for the following reason: when the guide field is strong, both $B_{z}\left(x_{0}\right)$ and $B_{z0}\left(x_{0}\right)$
are close to constant, with different small
variations to account for the force balance under different conditions, and
these differing small variations give rise to different derivatives for the two fields.
To obtain an approximation for $B_{z}'\left(x_{0}\right)$, we 
take the derivative of Eq.~(\ref{eq:constraint}) with respect
to $x_{0}$, yielding 
\begin{equation}
\frac{df}{dx_{0}}=-2B_{z}B_{z}'-2p'\simeq-2B_{z0}B_{z}'-2p_{0}'.\label{eq:df_dx}
\end{equation}
Hence, 
\begin{equation}
B_{z}'\simeq\frac{-p_{0}'}{B_{z0}}-\frac{1}{2B_{z0}}\frac{df}{dx_{0}}.\label{eq:Bz_prime}
\end{equation}

Using Eq.~(\ref{eq:Bz_prime}) in Eq.~(\ref{eq:Jp}) we can obtain
the poloidal current density 
\begin{equation}
\boldsymbol{J}_{p}=\frac{dB_{z}}{d\psi}\nabla\psi\times\boldsymbol{\hat{z}}=-\left(\frac{d\psi}{dx_{0}}\right)^{-1}\frac{dB_{z}}{dx_{0}}\boldsymbol{B}_{p}\simeq\frac{1}{x_{0}}\left[p_{0}'+\frac{1}{2}\frac{df}{dx_{0}}\right]\frac{\boldsymbol{B}_{p}}{B_{z0}}.\label{eq:J_poloidal}
\end{equation}
In the vicinity of $x_{0}=0$, $f\sim\left|x_{0}\right|^{8/3}$ and
$df/dx_{0}\sim\left|x_{0}\right|^{5/3}$. Assuming $p_{0}'\neq0$,
the pressure gradient term dominates and the poloidal current density
diverges as $J_{p}\sim p_{0}'/x_{0}$.

Next, we calculate the the out-of-plane current density $J_{z}=\boldsymbol{\hat{z}}\cdot\nabla\times\boldsymbol{B}_{p}$.
The dominant component of $\boldsymbol{B}_{p}$ is $B_{y}\simeq\text{sgn}\left(x_{0}\right)\sqrt{f(x_{0})+g(y)}$.
Therefore, the out-of-plane current density
\begin{equation}
J_{z}\simeq\left(\frac{\partial B_{y}}{\partial x}\right)_{y}=\left(\frac{\partial x}{\partial x_{0}}\right)^{-1}\frac{\partial B_{y}}{\partial x_{0}}\simeq\frac{\sqrt{f+g}}{\left|x_{0}\right|}\frac{\text{sgn}\left(x_{0}\right)}{2\sqrt{f+g}}\frac{df}{dx_{0}}=\frac{1}{2x_{0}}\frac{df}{dx_{0}},\label{eq:Jz-1}
\end{equation}
which is not affected by the pressure. Using the asymptotic behavior
(\ref{eq:f_0}), the leading order behavior of $J_{z}$ near $x_{0}=0$
is
\begin{equation}
J_{z}\simeq\frac{4}{3}c_{f}^{8/3}\left|x_{0}\right|{}^{2/3}.\label{eq:Jz_near0}
\end{equation}
Hence, the out-of-plane component $J_{z}\to0$ as $x_{0}\to0$. 

Now, we put our results in the context of the conventional treatment of the 
Pfirsch--Schl\"uter current density.  For a solution of the GS equation, the perpendicular component of the current density is
\begin{eqnarray}
\boldsymbol{J}_{\perp}&=&\frac{\boldsymbol{B}\times\nabla p}{B^{2}}=\frac{1}{B^{2}}\left(B_{z}\boldsymbol{\hat{z}}+\boldsymbol{\hat{z}}\times\nabla\psi\right)\times\nabla p \nonumber \\
&=&\frac{1}{B^{2}}\frac{dp}{d\psi}\left(B_{z}\boldsymbol{B}_{p}-B_{p}^{2}\boldsymbol{\hat{z}}\right), \label{eq:J_perp_2}
\end{eqnarray}
and the divergence of $\boldsymbol{J}_{\perp}$ is
\begin{eqnarray}
\nabla\cdot\boldsymbol{J}_{\perp}&=&\left(\boldsymbol{B}\times\nabla p\right)\cdot\nabla\frac{1}{B^{2}}=\frac{dp}{d\psi}\left(B_{z}\boldsymbol{B}_{p}-B_{p}^{2}\boldsymbol{\hat{z}}\right)\cdot\nabla\frac{1}{B^{2}} \nonumber \\
&=&\boldsymbol{B}_{p}\cdot\nabla\left(\frac{B_{z}}{B^{2}}\frac{dp}{d\psi}\right).\label{eq:div_J_perp}
\end{eqnarray}
Here, we have used that $z$ is the direction of symmetry and that both $p$ and $B_z$ are flux functions. 

The parallel current density is given by
\begin{eqnarray}
J_{\parallel}&=&\frac{J_{z}B_{z}+\boldsymbol{J}_{p}\cdot\boldsymbol{B}_{p}}{B}=\frac{1}{B}\left(B_z\nabla^2\psi -\frac{dB_{z}}{d\psi}B_{p}^{2} \right) \nonumber \\
&=&\frac{1}{B}\left(B_z\nabla^2\psi - \frac{dB_z}{d\psi}\left( B^2 -B_z^2\right) \right)  \nonumber \\
&=&-B\frac{dB_{z}}{d\psi}-\frac{B_{z}}{B}\frac{dp}{d\psi}.
\label{eq:J_parallel_1}
\end{eqnarray}
Here, in the last step we have applied the GS equation (\ref{eq:GS}). From Eqs.~(\ref{eq:div_J_perp}) and (\ref{eq:J_parallel_1}), we see that the magnetic differential equation~(\ref{eq:magnetic_de}) is indeed satisfied:
\begin{equation}
\boldsymbol{B}\cdot\nabla\frac{J_{\parallel}}{B}=-\boldsymbol{B}_{p}\cdot\nabla\left(\frac{B_{z}}{B^{2}}\frac{dp}{d\psi}\right)=-\nabla\cdot\boldsymbol{J}_{\perp}. \label{eq:mde}   
\end{equation}

For the HKT problem, the parallel and the perpendicular components of the current density can be obtained from Eqs.~(\ref{eq:J_parallel_1}) and (\ref{eq:J_perp_2}) as 
\begin{equation}
J_{\parallel}=-\frac{B}{x_0}\frac{dB_{z}}{dx_0}-\frac{B_{z}}{x_0B}\frac{dp}{dx_0}\simeq \frac{B}{B_z}\left( \frac{1}{2x_0}\frac{df}{dx_0} + \frac{p'}{x_0}\frac{B_p^2}{B^2}\right)
\label{eq:J_parallel}
\end{equation}
and
\begin{equation}
\boldsymbol{J}_{\perp}=\frac{p'}{x_0 B^{2}}\left(B_{z}\boldsymbol{B}_{p}-B_{p}^{2}\boldsymbol{\hat{z}}\right ). \label{eq:J_perp-1}   
\end{equation}
Alternatively, we can also obtain $J_\parallel$ and $\boldsymbol{J}_{\perp}$ from Eqs.~(\ref{eq:J_poloidal}) and (\ref{eq:Jz-1}) by making appropriate projections. Note that $J_\parallel \propto B_p^2/B^2$ and $J_\perp \propto B_p/B$; therefore, in the strong-guide-field limit the perpendicular component $J_\perp$ dominates.

Equations  (\ref{eq:J_parallel}) and (\ref{eq:J_perp-1}) show that the Pfirsch--Schl\"uter current density $J_{\parallel}$ and the diamagnetic current density $\boldsymbol{J}_{\perp}$ both diverge as $p'/x_{0}$.  If we assume that the system size along the $z$ direction is the same as the size $L$ along the $y$ direction, the equivalent ``rotational'' transform is given by
\begin{equation}
\iota\left(x_{0}\right)=\frac{B_{y0}}{B_{z0}}=\frac{x_{0}}{B_{z0}}.\label{eq:iota}
\end{equation}
Now, if we use the symbol $x$ to represent the rotational transform $\iota$ for the time being, the Pfirsch--Schl\"uter current density does have a $1/x$-type singularity; however, the underlying reason for its divergence is not what is commonly assumed.

In the conventional picture, the $1/x$-type singularity arises because the magnetic differential operator $\boldsymbol{B}\cdot \nabla$ is thought to become non-invertible at rational surfaces, as suggested by the expressions (\ref{eq:magnetic_de}) and (\ref{eq:magnetic_de1}) using a straight-field-line coordinate system. It is commonly assumed that $\nabla \cdot \boldsymbol{J}_\perp$ and the straight-field-line coordinate system are well-behaved; therefore, the $1/x$-type singularity follows from Eq.~(\ref{eq:J_parallel-1}). In contrast, our results show that the diamagnetic current density $\boldsymbol{J}_\perp$ diverges as $\sim p'/x_0$ and likewise, from Eq.~(\ref{eq:div_J_perp}), that $\nabla \cdot \boldsymbol{J}_\perp$ diverges as $\sim p'/x_0$ as well.  A crucial point is that because of the formation of the Dirac $\delta$-function current singularity, the poloidal magnetic field $\boldsymbol{B}_p$ does not go to zero as at the resonant surface; instead, $\boldsymbol{B}_p \to \pm \sqrt{g(y)} \boldsymbol{\hat{y}}$ as $x_0 \to \pm 0$. Consequently, the magnetic differential operator $\boldsymbol{B}\cdot \nabla$ is invertible, and the magnetic differential equation (\ref{eq:magnetic_de}) implies that the parallel current density $J_\parallel$ diverges in the same way as $\nabla \cdot \boldsymbol{J}_\perp$.\footnote{Strictly speaking, it may be more appropriate to describe the operator $\boldsymbol{B}\cdot \nabla$ as conditionally invertible at $x_0=\pm 0$. Because the function $g$ goes to zero at $y=0$ and $L$, the equation $\pm \sqrt{g(y)} d\varrho/dy = \varsigma$ is solvable only when $\varsigma$ approaches zero sufficiently fast as $y$ approaches $0$ and $L$ such that the function $\varsigma/\sqrt{g}$ is integrable. This condition is satisfied for $-\nabla \cdot \boldsymbol{J}_\perp$ on the right-hand side of the magnetic differential equation (\ref{eq:magnetic_de}).}

How do we reconcile the fact that the operator $\boldsymbol{B}\cdot \nabla$ is invertible at the resonant surface with its non-invertible appearance in a straight-field-line coordinate system? To answer this question, it is instructive to construct such a coordinate system. Because the initial magnetic field lines are straight, the Cartesian coordinates are a straight-field-line coordinate system. Now, if we take a Lagrangian perspective by describing the final state in terms of the mapping from the initial positions
$\boldsymbol{x}_0=(x_0, y_0, z_0)$ of fluid elements to their final positions $\boldsymbol{x}=(x,y,z)$, the initial coordinates $(x_0, y_0, z_0)$ are also a
straight-field-line coordinate system for the final state due to the ideal MHD frozen-in constraint. In the Lagrangian perspective, the magnetic field at $\boldsymbol{x}$ is determined by the initial
magnetic field $\boldsymbol{B}_{0}$ at $\boldsymbol{x}_{0}$ and the mapping $\boldsymbol{x}\left(\boldsymbol{x}_{0}\right)$ via the relation \citep{Newcomb1962,ZhouQBB2014}
\begin{equation}
\boldsymbol{B}=\mathcal{J}^{-1}\boldsymbol{B}_{0}\cdot\frac{\partial\boldsymbol{x}}{\partial\boldsymbol{x}_{0}},\label{eq:B_lagrangian}
\end{equation}
where $\mathcal{J}=\det\left(\partial\boldsymbol{x}/\partial\boldsymbol{x}_{0}\right)$
is the Jacobian of the mapping. 

\begin{figure}
\begin{center}
    \includegraphics[width=0.8\columnwidth]{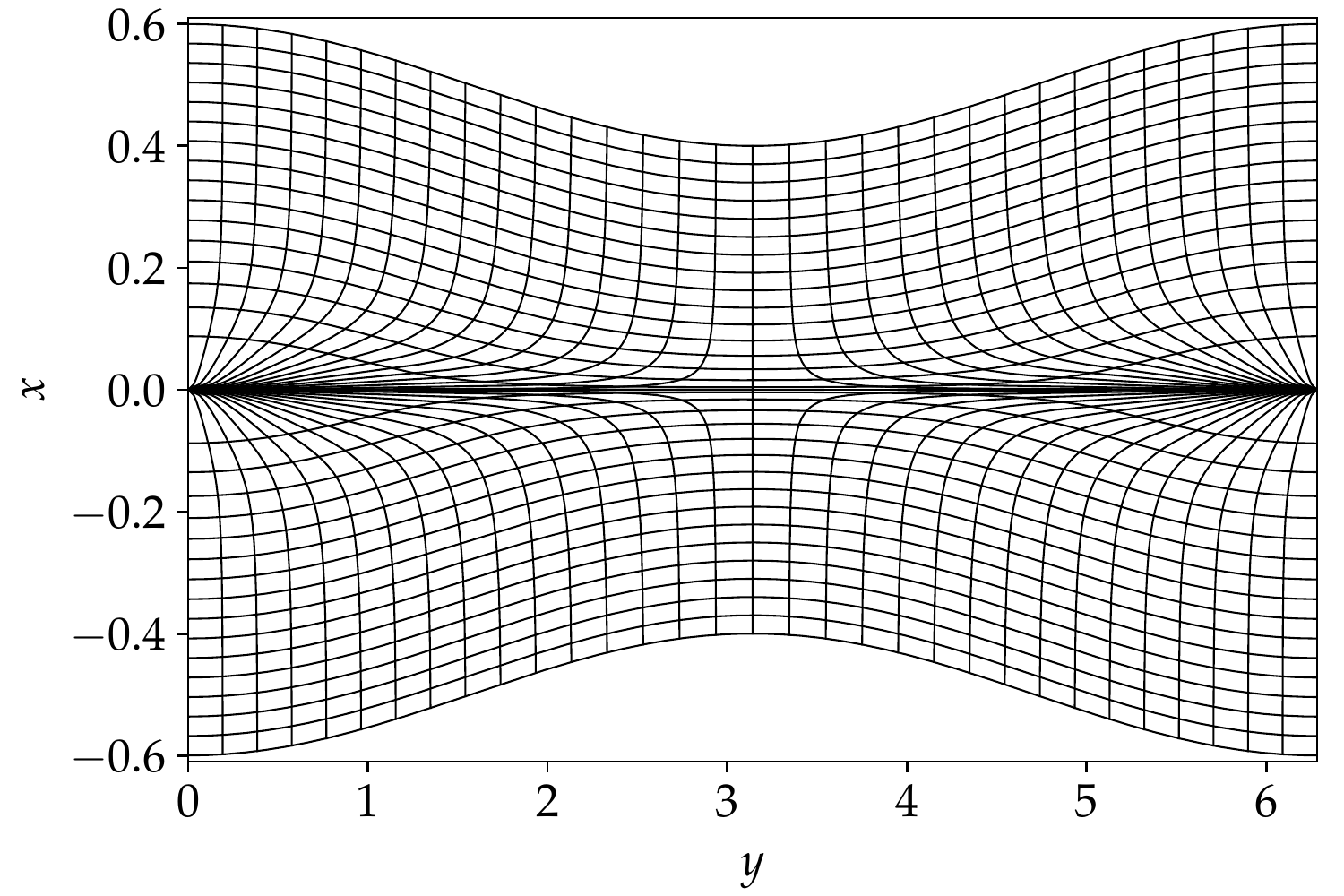}
\end{center}
\caption{A straight-field-line coordinate system for the HKT problem with $L=2\pi$, $a=0.5$, and $\delta=0.1$. The coordinate system becomes ill-behaved at the resonant surface. \label{fig:Straight-field-line}}
\end{figure}

Although our GS formulation is not fully Lagrangian, we
can reconstruct the Lagrangian mapping of fluid elements from
the initial to the final state once the solution is obtained. In this 2D problem, the fluid motion is limited to an $x$-$y$ plane. So, the $z$ coordinates of the initial and final states are identical, $z=z_0$. In an $x$-$y$ plane, for
each fluid element labeled by $\left(x_{0},y\right)$ in the final
state, we need to find its initial position $\left(x_{0},y_{0}\right)$.
This ``inverse'' Lagrangian mapping can be expressed as a function
$y_{0}\left(x_{0},y\right)$. From the conservation of magnetic flux
through an infinitesimal fluid element 
\begin{equation}
B_{z0}\left(x_{0}\right)dx_{0}\left[\frac{\partial y_{0}}{\partial y}dy\right]=B_{z}\left(x_{0}\right)\left[\frac{\partial x}{\partial x_{0}}dx_{0}\right]dy\label{eq:conservation_flux}
\end{equation}
and using Eq.~(\ref{eq:Bz}) to relate $B_{z0}$ and $B_{z}$, we
can calculate 
\begin{equation}
\frac{\partial y_{0}}{\partial y}=\frac{\partial x/\partial x_{0}}{\left\langle \partial x/\partial x_{0}\right\rangle }\label{eq:dy0_dy}
\end{equation}
and integrate it along each constant-$x_{0}$ contour to obtain $y_{0}(x_{0},y)$. Using the RDR solution (\ref{eq:RDR}) in Eq.~(\ref{eq:dy0_dy}) yields 
\begin{equation}
    \frac{\partial y_{0}}{\partial y}=\frac{\left| x_0 \right|}{\sqrt{f(x_0)+g(y)}}.   
    \label{eq:dy0_dy_RDR}
\end{equation}
In the limit of $x_0\to 0$, $\partial y_0/\partial y \to 0$ everywhere except at $y=0$ and $y=L$, where $\partial y_0/\partial y$ diverges as $\sim \left| x_0 \right|^{-1/3}$.
Therefore, the straight-field-line coordinate system $(x_0, y_0, z_0)$ becomes ill-behaved at the resonant surface, and that explains why the invertible operator $\boldsymbol{B}\cdot \nabla$ appears to be non-invertible in such a coordinate system. Figure \ref{fig:Straight-field-line} shows a constant-$z_0$ slice of the  straight-field-line coordinate $(x_0, y_0, z_0)$ for the HKT problem with $L=2\pi$, $a=0.5$, and $\delta=0.1$, where the solid lines are constant-$x_0$ or constant-$y_0$ coordinate curves. The singular nature of the coordinate system at the resonant surface is evident, as all the constant-$y_0$ coordinate curves converge towards $y=0$ or $y=L$.

Now, the critical question is: Does the $1/x$-type current singularity
lead to an infinite current?
Fortunately, the answer is no, and the reason can be attributed to
the Dirac $\delta$-function singularity that simultaneously appears
at the resonant surface. The finite tangential discontinuity $\left\llbracket B_{y}\right\rrbracket _{x_{0}=0} = 2\sqrt{g(y)}$, where $\left\llbracket \cdot\right\rrbracket $ denotes the jump across an interface, arises from a continuous initial magnetic field through squeezing the space between flux surfaces \citep{ZhouHQB2016}. As we can infer from the RDR solution (\ref{eq:RDR}), for flux surfaces sufficiently close to the resonant surface such that the condition
\begin{equation}
f\left(x_{0}\right)\ll g(y)\label{eq:condition}
\end{equation}
is satisfied, the mapping from the flux surface label $x_{0}$ to
the physical distance $s$ is quadratic:
\begin{equation}
s=\left|x\left(x_{0},y\right)\right|\simeq\frac{x_{0}^{2}}{\sqrt{g(y)}}.\label{eq:RDR-1}
\end{equation}
Because $f(x_0)\sim \left|x_0\right|^{8/3}$ and $g(y)\simeq\left(4c_{1}^{2}k^{2}/3\right)\sin^{8}(ky/2)$, where $k=2\pi/L$, the condition (\ref{eq:condition}) eventually will be satisfied with a sufficiently small $x_{0}$ for all $y$ except at $y=0$ and $y=L$. As such, the mapping eventually will become quadratic, $s\sim x_{0}^{2}$, although the transition to the quadratic mapping  will occur at different $x_{0}$ for different $y$. Because of the quadratic mapping, the poloidal current density $J_{p}\sim1/x_{0}$ becomes $J_{p}\sim1/\sqrt{s}$. And since $\int_{0}^{s}ds'/\sqrt{s'}$ is integrable, the total current does not diverge. 

What about at $y=0$ and $y=L$ where the mapping is not quadratic? Because $g=0$ at those locations and $f\sim \left|x_{0}\right|^{8/3}$, Eqs.~(\ref{eq:J_poloidal}) gives $J_{p}\sim \left|x_{0}\right|^{1/3}$, which does not diverge. Moreover, to compensate for the strong squeezing of the quadratic mapping, the flux surfaces in the downstream regions near $y=0$ and $y=L$ have to bulge out. Consequently, the scaling at $y=0$ and $y=L$ becomes $s\sim \left|x_{0}\right|^{2/3}$; see Ref.~\citep{HuangHLZB2022}. 

Intuitively, the $J\sim1/\sqrt{s}$ scaling can be understood as follows.
The quadratic mapping $s\sim x_{0}^{2}$ causes a steepening of the
pressure gradient:
\begin{equation}
\frac{\partial p}{\partial x}=\left(\frac{\partial x}{\partial x_{0}}\right)^{-1}\frac{\partial p}{\partial x_{0}}\sim\frac{1}{x_{0}}\frac{\partial p}{\partial x_{0}}\sim\frac{1}{\sqrt{s}}\frac{\partial p}{\partial x_{0}}.\label{eq:p_steepening}
\end{equation}
Because the pressure gradient is balanced by the $\boldsymbol{J}\times\boldsymbol{B}$
force for an MHD equilibrium, the perpendicular current density $J_\perp \sim\left|\nabla p\right|\sim1/\sqrt{s}$, and $\nabla \cdot \boldsymbol{J}_\perp$ diverges in the same manner. Finally, the parallel current density satisfies the magnetic differential equation (\ref{eq:magnetic_de}), and the operator $\boldsymbol{B} \cdot \nabla$ is invertible; therefore, $J_\parallel \sim 1/\sqrt{s}$ as well.

\section{Numerical Verification of the Analytic Solutions\label{sec:Numerical-Solutions}}

\begin{figure}
\includegraphics{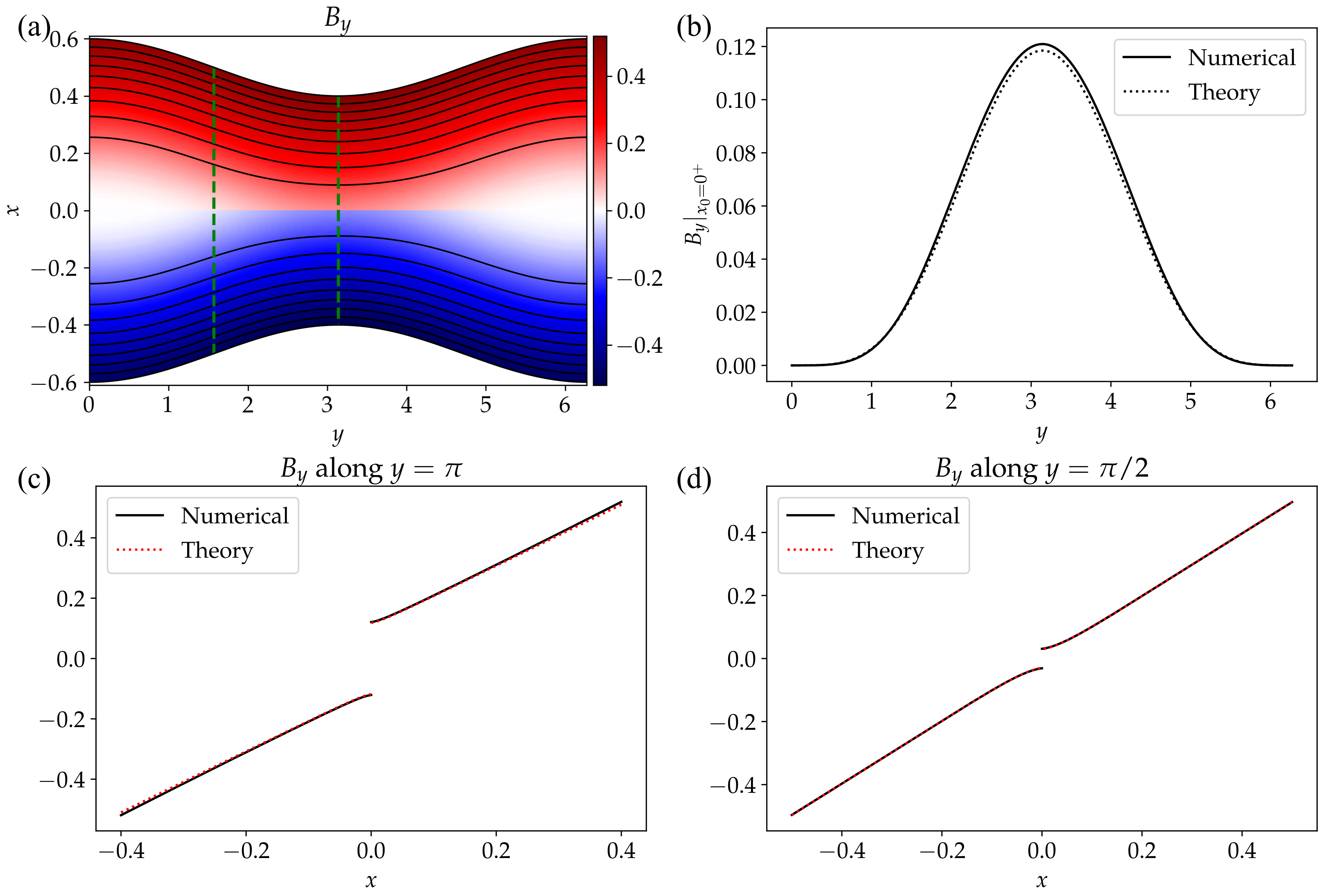}

\caption{Magnetic field component $B_{y}$ (a) in the entire domain; (b) along
$x_{0}=0^{+}$; (c) along $y=\pi$; (d) along $y=\pi/2$. Black solid lines in Panel (a) are flux surfaces, and vertical
dashed lines indicate the locations of the one-dimensional
cuts of Panels (c) and (d). Black solid lines in Panels (b\textendash d) are numerical results, and the red dotted lines are the predictions of the analytic theory.\label{fig:By}}
\end{figure}

We now compare the analytic solutions in  Section \ref{sec:Grad-Shafranov-solutions} with
numerical solutions using a GS solver. The GS solver has been extensively
tested, showing good agreement with the solutions of a fully Lagrangian
solver \citep{ZhouHQB2016} and the SPEC code \citep{HuangHLZB2022}
for the HKT problem with $p=0$. Here, we consider an initial equilibrium
with a linear pressure profile

\begin{equation}
p_{0}=\bar{p_{0}}+rx_{0}\label{eq:pressure0}
\end{equation}
and a magnetic field 

\begin{equation}
\boldsymbol{B}_{0}=x_{0}\boldsymbol{\hat{y}}+\sqrt{B_{0}^{2}-2rx_{0}-x_{0}^{2}}\boldsymbol{\hat{z}}.\label{eq:initial_B}
\end{equation}
The parameters in the following numerical calculations are: the domain
sizes $a=1/2$ and $L=2\pi$; the guide field $B_{0}=10$; the mean
pressure $\bar{p_{0}}=1$; the pressure gradient $r=1$; the perturbation
amplitude $\delta=0.1$; the adiabatic index $\gamma=5/3$. 

In our previous studies, we assumed a mirror symmetry across the mid-plane
and only solved the GS equation in half of the domain with $x_{0}\ge0$.
In this study, because of the asymmetric pressure profile across the resonant surface, we cannot impose
the mirror symmetry. We divide the domain into two regions separated
by the flux surface labeled by $x_{0}=0$ and solve the GS equation
in each region by a descent method. Simultaneously with the iteration
in both regions, the $x_{0}=0$ flux surface is allowed to move until
the force-balance condition $\left\llbracket B^{2}/2+p\right\rrbracket =0$
is satisfied.

\begin{figure}
\includegraphics[width=1\columnwidth]{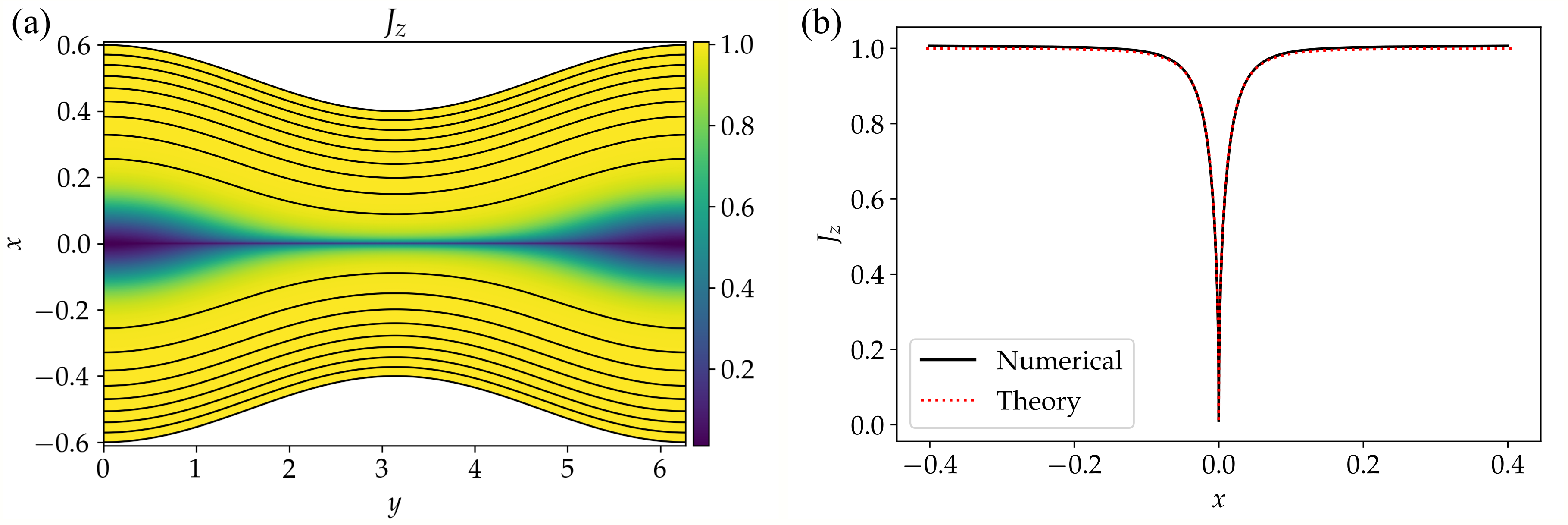}

\caption{(a) A numerical solution of $J_{z}$ in the entire domain; (b) 1D
cuts along $y=\pi$ in Panel (a). The black solid line in Panel (b)
is the numerical solution, and the red dotted line is the prediction
of the analytic theory.\label{fig:Jz}}
\end{figure}
\begin{figure}
\includegraphics[width=1\columnwidth]{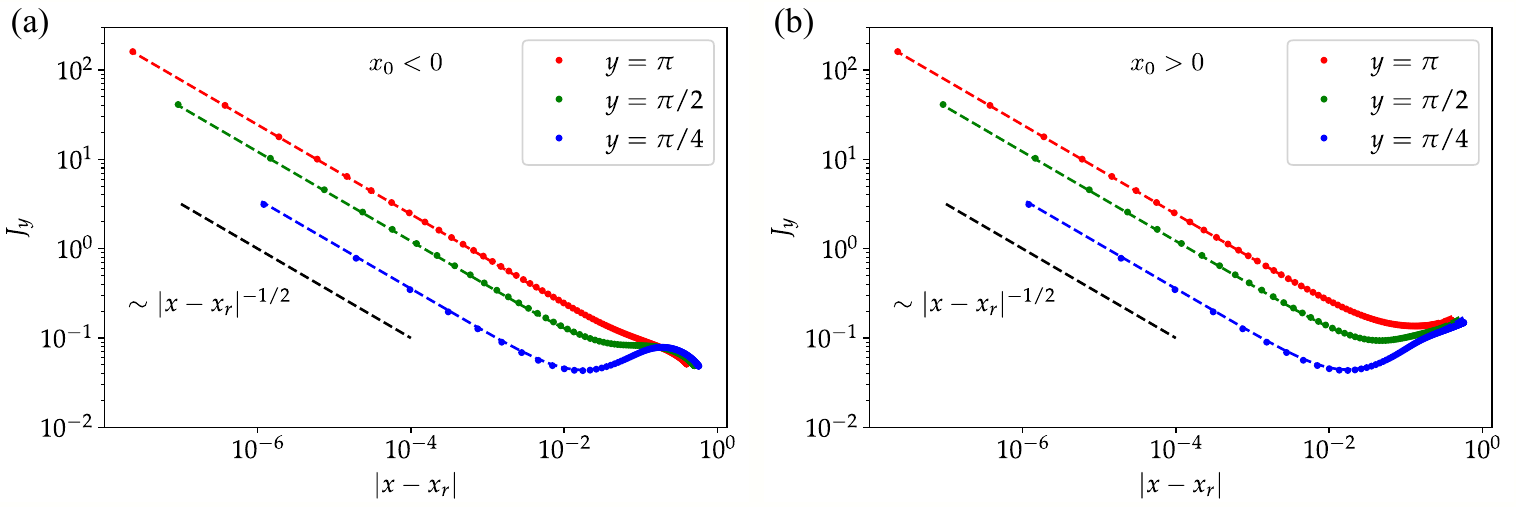}

\caption{Comparisons between the numerical solutions of $J_{y}$ and the theoretical
predictions along a few 1D cuts. Here, the solid dots represent the
numerical solutions and the dashed lines are theoretical predictions.
Panel (a) shows the region with $x_{0}<0$ and Panel (b) shows the
region with $x_{0}>0$. Here, the horizontal axis $\left|x-x_{r}\right|$
is the distance between the flux surface at $x$ and the resonant
surface at $x_{r}$. As predicted by the theory, $J_{y}$ diverges
as $\left|x-x_{r}\right|^{-1/2}$ near the resonant surface. \label{fig:Jy}}
\end{figure}
\begin{figure}
\includegraphics[width=1\columnwidth]{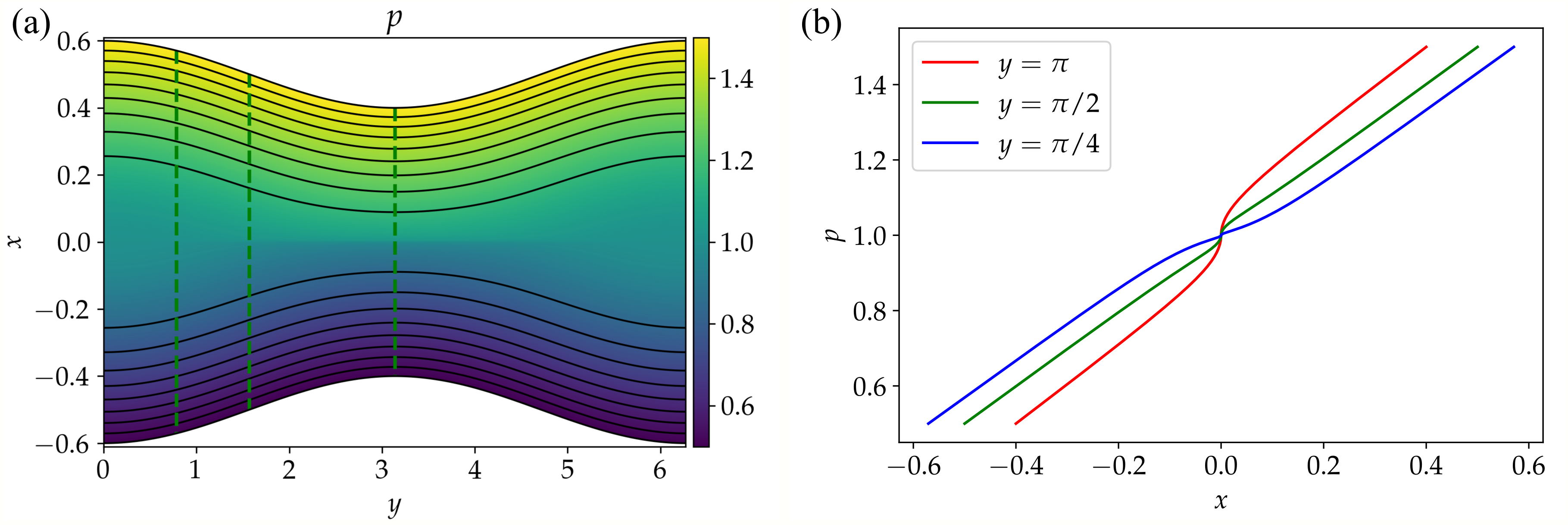}

\caption{(a) A numerical solution of the pressure $p$ in the entire domain;
(b) 1D cuts along the dashed lines in Panel (a).\label{fig:pressure}}
\end{figure}

Panel (a) of Figure \ref{fig:By} shows a numerical solution of the
\textbf{$B_{y}$ }component in the entire domain. Panels (b\textendash d)
show a few selected one-dimensional (1D) cuts of the numerical solution,
together with the corresponding analytical solution (\ref{eq:Inner_sol}).
The analytic and the numerical solutions are in good agreement. The
magnetic field is discontinuous across the resonant surface, corresponding
to a Dirac $\delta$-function current singularity.

Figure \ref{fig:Jz} compares the numerical solution of the $J_{z}$
component with the theoretical prediction using Eq.~(\ref{eq:Jz-1}).
Here, $J_{z}\to0$ at the resonant surface as the theory predicts.
Likewise, Figure \ref{fig:Jy} shows comparisons between the numerical
solutions of $J_{y}$ along selected 1D cuts and the theoretical predictions
using Eq.~(\ref{eq:J_poloidal}). In this figure, the horizontal
axis $\left|x-x_{r}\right|$ is the distance between the flux surface
at $x$ and the resonant surface at $x_{r}$. As predicted by the
theory, $J_{y}$ diverges as $\left|x-x_{r}\right|^{-1/2}$ near the
resonant surface. 

Finally, Figure \ref{fig:pressure} shows a numerical solution of
the pressure $p$ in the entire domain and along selected 1D cuts.
Here, the steepening of the pressure gradient due to the squeezing
of flux surfaces near the resonant surface is evident. Examining the
solutions in the vicinity of the resonant surface indicates that the
$\partial p/\partial x\sim\left|x-x_{r}\right|^{-1/2}$, as expected
from our analysis.

As a remark, although the analytic theory is derived under the strong-guide-field
limit, we find that the theoretical predictions remain close to numerical
results even for a moderate guide field such as $B_{0}=2$. 

\section{Discussions and Conclusion\label{sec:Discussions-and-Conclusion}}

In summary, we have derived an analytic theory for the pressure-gradient-driven current singularity near the resonant surface of the ideal Hahm\textendash Kulsrud\textendash Taylor
(HKT) problem and have validated the theory with numerical solutions. Our key finding is that although the current density diverges as $J\sim1/\Delta\iota$, where $\Delta\iota$ is the difference of the rotational transform relative to the resonant surface, this singularity does not lead to a divergent total current. The reason is that the distance $s$ between the resonant surface and an adjacent flux surface is not proportional to $\Delta\iota$. Because of the formation of a Dirac $\delta$-function current sheet at the resonant surface, the neighboring flux surfaces are strongly packed, and the distance $s\sim(\Delta\iota)^{2}$. Consequently, the current density $J \sim 1/\sqrt{s}$, which is integrable and the total current is finite.

Our analysis also finds that the Pfirsh--Schl\"uter current density $J_\parallel$ and the diamagnetic current density $\boldsymbol{J}_\perp$ both diverge as $1/\sqrt{s}$, where the diamagnetic current density is the dominant component. The diamagnetic current density diverges because the strong packing of flux surfaces near the resonant surface causes a steepening of the pressure gradient; therefore, $\boldsymbol{J}_\perp \sim dp/ds\sim1/\sqrt{s}$. Furthermore, solving the magnetic differential equation (\ref{eq:magnetic_de}) yields the parallel current density $J_\parallel$ diverging in the same manner as the perpendicular component $\boldsymbol{J}_\perp$. Notably, contrary to the conventional wisdom that the pressure needs to flatten at rational surfaces to have an integrable current density, here we show that  the current singularity remains integrable despite a steepening of the pressure gradient. 

Our analysis assumes that the initial pressure gradient $dp/dx_{0}$
does not diverge and is non-vanishing at $x_{0}=0$. More generally,
we may consider an initial condition with $J_{p}\sim dp/dx_{0}\sim\left|x_{0}\right|^{\alpha}$,
where $\alpha>-1$ such that the current density is integrable. After
the boundary perturbation, the current density $J\sim dp/ds\sim s^{(\alpha-1)/2}$.
Because the power index $(\alpha-1)/2>-1$ if $\alpha>-1$, the current
density is integrable. Furthermore, we may consider an initial magnetic
field $B_{y0}\sim\text{sgn}\left(x_{0}\right)\left|x_{0}\right|^{\nu}$.
After the boundary perturbation, the formation of a Dirac $\delta$-function current sheet requires the distance between flux surfaces to scale as $s\sim\left|x_{0}\right|^{\nu+1}$; consequently, the current density $J\sim dp/ds\sim s^{(\alpha-\nu)/(\nu+1)}$,
which is integrable. In all these cases, as long as the initial current
density is integrable, the current density in the perturbed state
is also integrable. 

Although our conclusion is obtained with a simple prototype problem,
a similar conclusion probably applies to more general magnetic
fields and potentially resolves the paradox of non-integrability of the Pfirsch--Schl\"uter
current density. If that is to be the case, then a weak
(i.e., non-smooth) ideal MHD equilibrium solution with a continuum
of nested flux surfaces, a continuous pressure distribution with non-vanishing pressure gradients on rational surfaces, and a
continuous rotational transform is not prohibited. Such an equilibrium
may be pathological, to quote Grad, but nonetheless mathematically
intriguing to contemplate. 

\ack{}{}

We thank Dr.~Greg Hammett for beneficial discussion. This research
was supported by the U.S. Department of Energy under Contract No.
DE-AC02-09CH11466 and by a grant from the Simons Foundation/SFARI
(560651, AB). Part of this work has been carried out within the framework
of the EUROfusion Consortium and has received funding from the Euratom
research and training programme 2014\textendash 2018 and 2019\textendash 2020
under Grant Agreement No. 633053. The views and opinions expressed
herein do not necessarily reflect those of the European Commission.
YZ was sponsored by Shanghai Pujiang Program under Grant No.
21PJ1408600. Part of the numerical calculations were performed with
computers at the National Energy Research Scientific Computing Center.

\bibliographystyle{unsrt}
\addcontentsline{toc}{section}{\refname}\bibliography{ref.bib}

\end{document}